# Operation and coupling of LH waves with the ITER-like wall at JET

K. K. Kirov[1], J. Mailloux[1], A. Ekedahl[2], V. Petrzilka[3], G. Arnoux[1], Yu. Baranov[1], M. Brix[1], M. Goniche[2], S. Jachmich[4], M.-L. Mayoral[1], J. Ongena[4], F. Rimini[1], M. Stamp[1] and JET EFDA Contributors*

*JET-EFDA, Culham Science Centre, Abingdon, OX14 3DB, UK*
*[1]EURATOM/CCFE Fusion Association, Culham Science Centre, Abingdon, OX14 3DB, UK*
*[2]CEA, IRFM, F-13108 Saint Paul-Lez-Durance, France*
*[3]Association EURATOM-IPP.CR, Institute of Plasma Physics AS CR, Za Slovankou 3, 182 21 Praha 8, Czech Republic*
*[4]Association "EURATOM - Belgian State", ERM, Brussels, Belgium*

**Abstract:** In this paper important aspects of Lower Hybrid (LH) operation with the ITER Like Wall (ILW) [1] at JET are reported. Impurity release during LH operation was investigated and it was found that there is no significant *Be* increase with LH power. Concentration of *W* was analysed in more detail and it was concluded that LH contributes negligibly to its increase. No cases of *W* accumulation in LH-only heating experiments were observed so far. LH wave coupling was studied and optimised to achieve the level of system performance similar to before ILW installation. Measurements by Li-beam were used to study systematic dependencies of the SOL density on the gas injection rate from a dedicated gas introduction module and the LH power and launcher position. Experimental results are supported by SOL transport modelling. Observations of arcs in front of the LH launcher and hotspots on magnetically connected sections of the vessel are reported. Overall, a relatively trouble-free operation of the LH system up to 2.5MW of coupled Radio Frequency (RF) power in L-mode plasma was achieved with no indication that the power cannot be increased further.

## INTRODUCTION

ITER is designed to operate with beryllium (*Be*) main vessel and tungsten (*W*) divertor tiles due to the high fuel retention found on the carbon (*C*) based machines [2]. These are relatively new materials to be used as Plasma Facing Components (PFC) and little is known about the expected plasma performance of such all-metallic wall devices. The necessity of

* See the Appendix of F. Romanelli et al., Proceedings of the 24th IAEA Fusion Energy Conference 2012, San Diego, US

understanding the physics of plasma surface interaction and exploring the operational space for ITER initiated a project on installation of a new ITER Like Wall (ILW) at JET [1], [3]. The new JET wall has outboard Poloidal Limiters (PLs) and inner guard limiters made of *Be*, while the thermal load bearing divertor consists of *W* coated *C* tiles and bulk *W* tile assemblies.

After the installation of the ILW a number of key operational issues were addressed including: reinstating the robust plasma breakdown; being able to perform stable recovery pulses; and implementing protection of the new wall from plasma exhaust and energetic particles. In order to avoid melting the *Be* limiters, some restrictions on the heating power were necessary. Further, to reduce sputtering yield from divertor *W* tiles, JET needed to operate at lower SOL temperature, which was achieved by use of large gas puff rates and hence operation at higher density.

Although currently not planned, a LH system is still under consideration for ITER [4], [5], [6]. While up-to-date experimental and theoretical studies are mainly focused on LH wave coupling and penetration issues and current drive (CD) efficiency at ITER-like densities [7], [8], [9], [10], [11], [12], [13], [14], [15], [16], [17] no LH system has ever been tested at SOL conditions relevant to the PFC envisaged for ITER. Being equipped with LH system and ILW, JET provides a unique opportunity to study the performance of this Radio Frequency (RF) heating scheme in conditions as close as possible to ITER in terms of PFC. Nevertheless the use of Multi-Junction (MJ) instead of Passive Active Multijunction (PAM) launcher requires more gas for good coupling, which together with the lower power to major radius factor of JET make SOL characteristics not directly extrapolable to those expected on ITER.

The use of the LH system at JET has been revised as a part of major review on the use of all heating systems in conditions with the new wall. The power was limited initially to 180kW per klystron and 2.5MW of total coupled RF power. The position of the LH launcher, which can be controlled during JET pulse, was constrained within certain limits at high power operations. An attempt to improve the protection against arcs and to minimise their occurrence and/or avoid hotspots has been made. The latter are localised hot areas on PFC and can cause damage to the wall while the arcs can lead to plasma disruptions.

A new protection system named Protection of ITER-like Wall (PIW) [18], based on a set of new dedicated Infra Red (IR) and visible cameras and pyrometers, was introduced to monitor critical PFC and to prevent the PFC from reaching temperatures high enough to cause melting and damage to the wall. The upgrade of the monitoring systems included installation and operation of a new camera with a dedicated view of the LH launcher. The diagnostic features a shared view for IR imaging, visible camera and four pyrometers. The IR camera and the pyrometers were used to study hotspots and possible overheated areas on launcher's frame and near poloidal limiter.

At the beginning of normal operation the focus of the JET program was more directed into issues related to fuel retention with the new wall, material studies regarding the new PFC, impurity accumulation and scenarios development. Priority was given to baseline and hybrid scenarios so LH was rarely used in the main program. As a result no LH pulses in H-mode were performed and results presented here are collected mainly from heating power conditioning sessions during the Restart period.

This paper gives an account on important aspects of the performance of the LH system with the ILW at JET for power levels up to 2.5MW and up to 5 seconds in duration. Initial observations were discussed in [19] and further developments are discussed below. A number of topics related to the new conditions associated with ILW are presented via comparison to the old *C* wall (CW). Throughout the paper the new data will be referred to as ILW conditions, new wall or new conditions while for the old *C* wall one of the following terms will be used: old conditions, old *C* wall or simply CW. If not explicitly noted in the text one should assume that all JET pulses prior to pulse number 79854 are in conditions with CW, whilst newer pulses are with ILW. The discussions addressed here are split into three major topics: (*i*) impurity release during LH operation; (*ii*) coupling issues in conditions with the new wall and (*iii*) observations and analysis of arcs and hotspots. Section 2 gives an account of the important changes to the LH system with installation of ILW. Section 3 discusses observations of impurities related to LH operation and ILW. Section 4 is dedicated to coupling studies and it also provides important results on SOL density measurements and modelling. Observations of arcs and hotspots are discussed in Section 5. The last Section is dedicated to conclusions and remarks regarding future operation of the system and on-going projects aiming at improvement of the arc detection system.

## LH SYSTEM AT JET

The LH system at JET is detailed in [20], [21] and more details, including a number of recent restrictions imposed on LH operations, are discussed in [22]. Here only an account of the system components which have been affected by the ILW installation will be given.

The LH launcher at JET has an active radiating part, so called multijunctions [23] or sometimes for simplicity called the "grill". The multijunctions are bound by a passive structure, referred to as a frame, which protrudes about 0.003m to 0.005m in front of the grill thus protecting it from interaction with hot plasma and energetic particles. A picture of the launcher front end, or so called mouth, with notations of the six rows and the launcher surroundings is shown in figure 1. Reflection Coefficients (RCs) are measured as a ratio of the reflected to the forward RF power averaged along each row, RC1 to RC6 for rows 1 to 6, and good coupling is defined to be RC≤0.08. If the electron density in front of the launcher is not sufficiently high the coupling of the RF wave deteriorates as a large amount of the RF power is reflected back to the waveguides. At JET there are a couple of ways to improve the coupling. The first method is to move the launcher closer to the denser plasma and the second is to use gas injection from the dedicated gas introduction module situated near the LH launcher, figure 1b. Two limiters are situated on each side of the launcher; the one on the left is normal Poloidal Limiter (PL) while the right-hand one is narrower PL (nPL). The smaller width of the nPL restricts its load bearing capacity so it is aligned 0.005m behind the normal PLs. The distance between the LH launcher and plasma is measured by two quantities: launcher to nPL distance, $l_{pos}$, and separatrix to PL clearance at midplane, $r_{ROG}$. Negative values of $l_{pos}$ mean the launcher is behind the nPL hence the real distance between the separatrix and launcher at midplane is the sum $r_{ROG}$+0.005-$l_{pos}$.

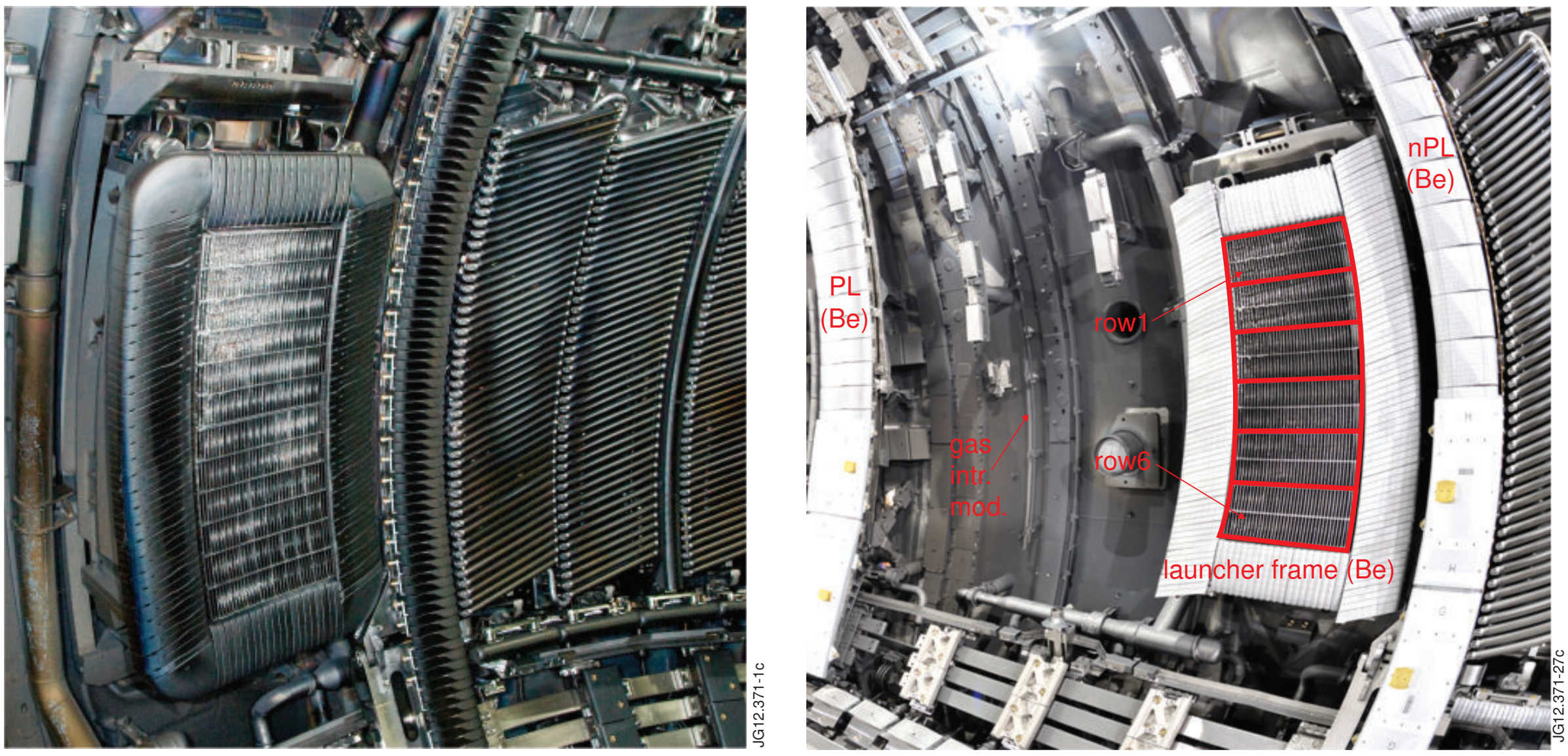

a) b)

**Figure 1:** The LH launcher as photographed from the left side in 2007 (a). A picture of the launcher taken from right side in 2011 after its frame was replaced by Be one (b). Notations of rows 1 and 6 are provided as well as launcher frame, gas introduction module, poloidal limiter (PL) and narrow poloidal limiter (nPL) surrounding the launcher.

During the Shutdown in 2010 a couple of modifications were made to the launcher. While all the sections of the old frame were made of *C* or CFC, figure 1a, the new tiles were manufactured from *Be*, figure 1b, in order to comply with the requirement for an all-*Be* PFC. The grill is made of stainless steel and this has not been changed. Although the new frame looks different from the old one, especially the corner sections, the geometry of the sides was kept essentially unchanged. The same applies for all the limiters surrounding the launcher meaning that, from the geometrical point of view, the scrape-off lengths are the same as with the old wall. The dedicated gas introduction module has not been changed.

## IMPURITIES RELATED TO ILW

Impurity release during LH operation was investigated using the available spectroscopic diagnostics at JET. Surface melting of PFC and PLs can be caused by excessive heat loads by fast electrons in SOL generated parasitically by LH waves [39]. The intensity of *Be* line radiation has been analysed in order to identify a possible link between *Be* influx and LH power. This can be interpreted as an indication of PFC being heated as a result of harmful LH wave – SOL plasma interaction. It was found [19], [24] that for up to 2.5MW there is no consistent increase in *Be* in LH-only experiments.

Tungsten generation and accumulation with application of auxiliary heating power, NBI and ICRH, of the order of 3.5MW was observed at JET [25]. Although for different heating schemes the underlying physical processes related to impurity generation and transport are of different nature, this prompted a similar investigation regarding the LH power. The concentration of $W$, $C_W = n_W/n_e$, is studied in more detail using a dedicated analysis code [26], [27], which collects data from available spectrometers and bolometers and uses plasma profiles from various diagnostics to assess the concentration of $W$. The increase in $W$ concentration during LH is found negligible [24]. Time traces of a typical LH conditioning pulse with magnetic field, plasma current, density, gas puff rate and an estimate of $C_W$ levels are shown in figure 2.

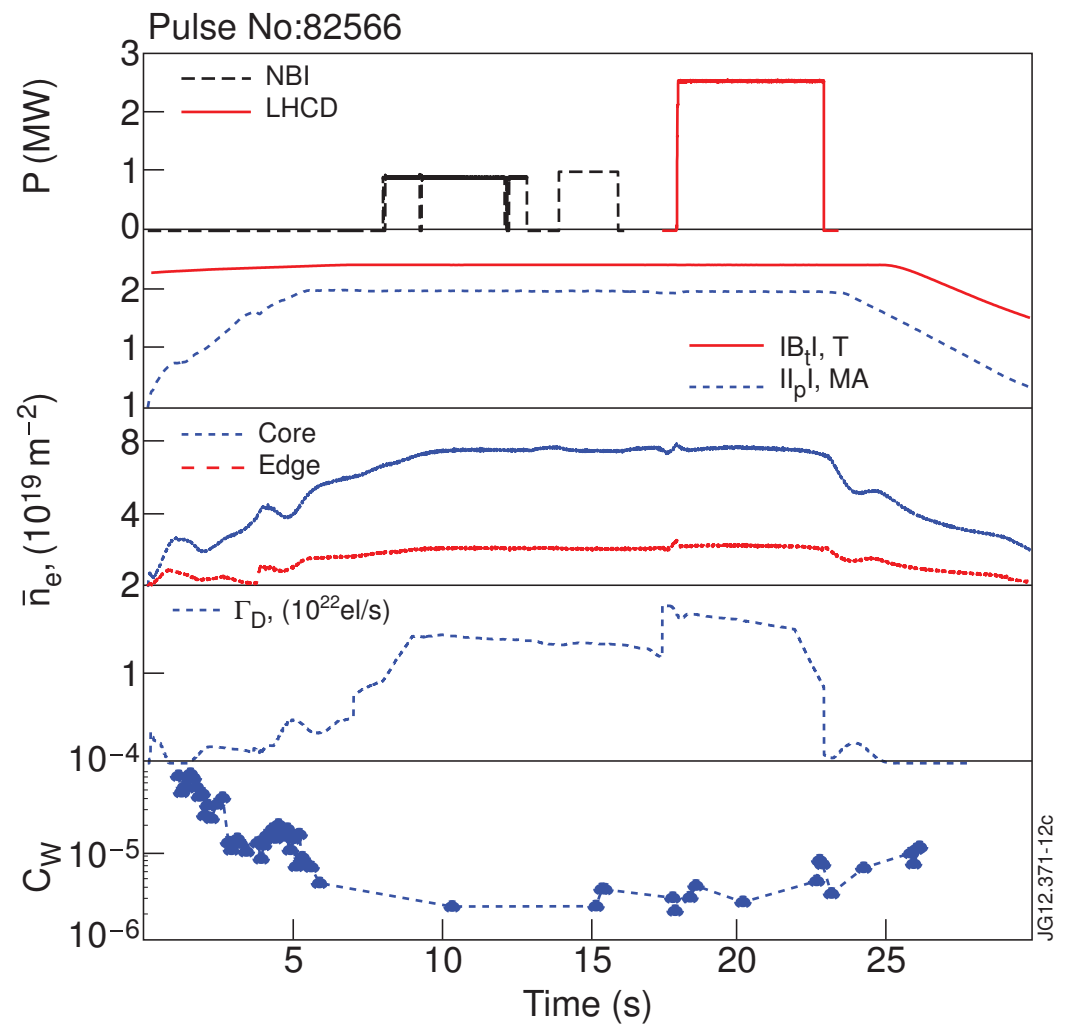

**Figure 2:** Time traces of a typical LH conditioning pulse showing from top to bottom LH and NBI power, $B_t$, $I_p$, core and edge density, total gas puff rates and $C_W$, which is equal to the W density normalised to local electron density.

It should be noted that during LHCD the concentration of $W$ is of the same order as during low-power (≈1MW) NBI heating and $C_W$ definitely does not increase more than twice compared to OH phase. The analysis has also shown that no persistent $W$ accumulation event has been detected during LH-only pulses.

## COUPLING STUDIES

During the initial phase of LH system operation after completing ILW installation a priority was given to bringing the system to level of performance achieved with the old wall. Therefore, not all parameters which could possibly affect LH wave coupling were investigated. For instance, the impact of the configuration was not fully studied as most of the LH related experiments were performed at low triangularity shape as shown in figure 3b. The only two adjustable parameters available for optimising the LH performance were the gas rate from dedicated gas injection valve and launcher position.

### LH wave coupling in conditions with the new wall.

Two main factors were expected to have a significant effect on the LH wave coupling with the ILW, compared to the old wall conditions. Due to the new metallic wall one would expect different recycling and impurities thus different SOL conditions. Also, in order to decrease the SOL temperature, thus reducing the sputtering yield, all operations with the ILW at JET used large gas puff rates. These issues were expected to have an impact on the LH performance so time was dedicated to studying their contributions in the new conditions of ILW.

Initial observations indicated that with launcher retracted behind nPL, i.e. for $l_{pos} < 0$m, RCs are large and gas injection of $D_2$ from dedicated gas valve is needed in order to maintain good coupling conditions.

A more detailed comparison of the coupling conditions was carried out for $l_{pos} \geq 0$m after cross-checking the averaged RCs for pulses in which SOL conditions and plasma parameters are similar. In figure 3 an example of two pulses, one with ILW and $l_{pos} \approx +0.001$m and the other with CW with $l_{pos} \approx +0.006$m and with similar total gas injection rates, configurations (figure 3b) and plasma-launcher distance are shown. As the launcher was placed closer to the plasma gas injection from dedicated valve was not used in these two cases. After comparing the coupling on individual rows (figure 3a bottom graph) it was found that the RCs on the top and at the middle of the launcher are approximately the same confirming the conclusion that for this part of the grill coupling is not affected. The bottom two rows, RC5&6, however, measure higher RCs with ILW compared to CW. It is possible that this is a result of slightly larger $l_{pos}$ in the ILW case, which is expected to affect mostly the bottom part of the launcher as the plasma-launcher distance is largest for rows 5&6 due to non-perfect match of the separatrix and the grill's shape (see figure 3b). Adding gas puff from a dedicated gas valve, a well known remedy to lower the RCs [28], [29], [30], [31] improves the coupling on the bottom rows. Although the example shown in figure 3 represents significant part of LH related experiments before ILW installation, it is unclear whether in different conditions, e.g. different configurations and launcher behind nPL, this conclusion will still hold. A full-scale analysis would require similar comparison for the whole range of parameters in which LH is used, including cases at high triangularity and with launcher in the shadow of nPL.

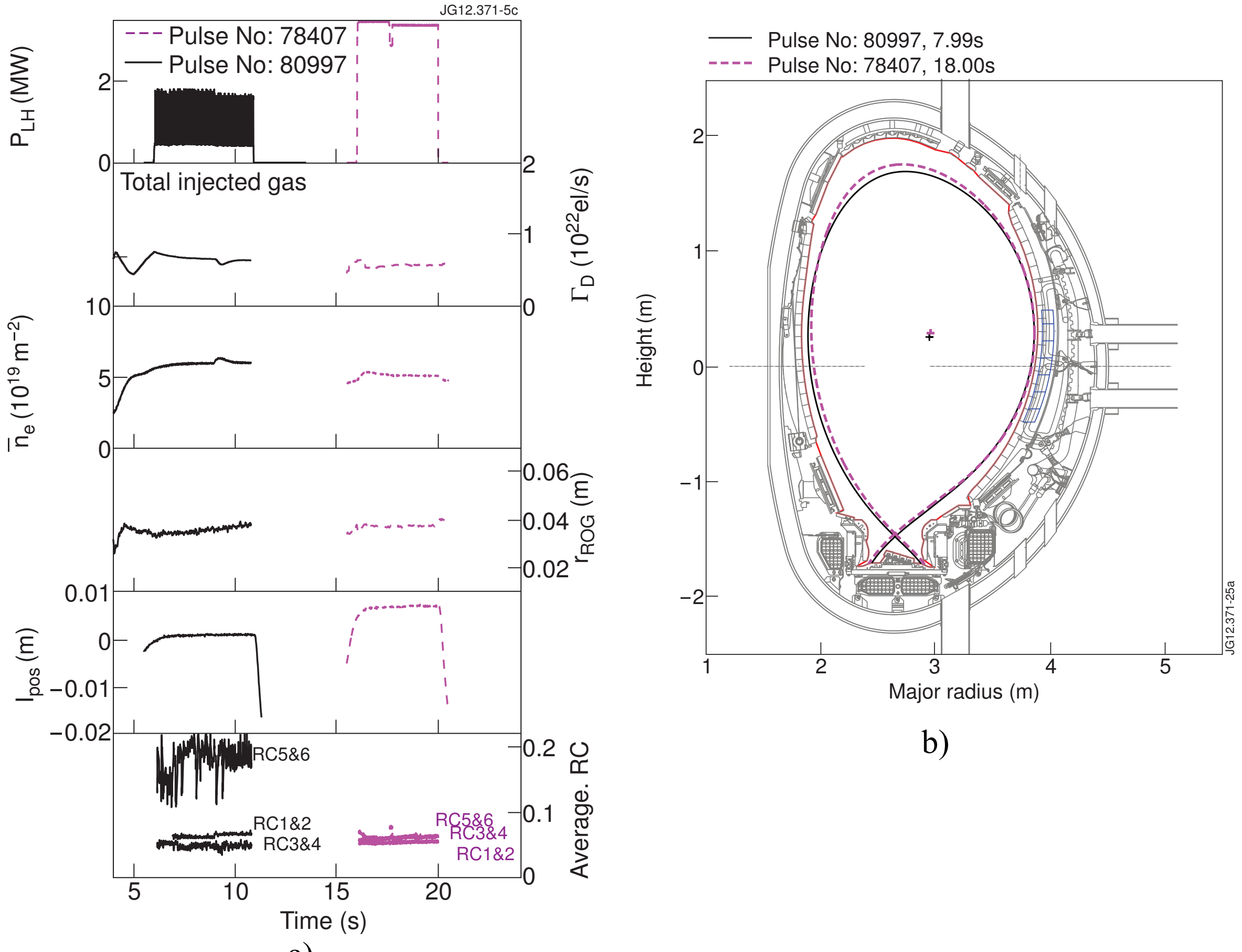

**Figure 3:** LH wave coupling with C wall and ILW conditions. Time traces of two similar pulses with no gas from dedicated gas injection valve, #78407 (2.2T/2MA, magenta) and #80997 (2.28T/2MA, black), are compared in a). From top to bottom shown are the LH power, total gas injection rate, line integrated density, plasma-limiter clearance, $r_{ROG}$, limiter launcher distance, $l_{pos}$, and averaged RCs on rows 1&2, 3&4, and 5&6 are shown. Plasma shapes for the two cases are given in b).

## Scan of gas injection rates from dedicated gas valve.

In the process of optimising the LH system performance, the gas injection rates were scanned and it was found that coupling improves on all rows with gas injection rate.

An example is shown in figure 4a, where RCs of rows 1 to 6, RC1 to RC6, are shown for four different values of the gas injection rates, namely 0, $2\times10^{21}$, $4\times10^{21}$, $6\times10^{21}$ el/s. Power waveforms, gas puff rates and line-integrated electron density are also shown in figure 4b. For the middle rows, 3&4, with RC3 and RC4, the coupling is good, RCs<0.08, even without gas. Higher rates of injection further reduce the RCs on these rows. The top two rows, 1&2, need rates of about $4\times10^{21}$el/s to achieve a reasonable coupling and to stop being tripped by the protection system, based on the imbalance in the reflected RF power. In contrast, the bottom rows, 5&6, show bad coupling, RCs>0.12, for lower gas rates and good coupling is only achieved at maximum gas injection rate used in the experiments, $\approx6\times10^{21}$el/s.

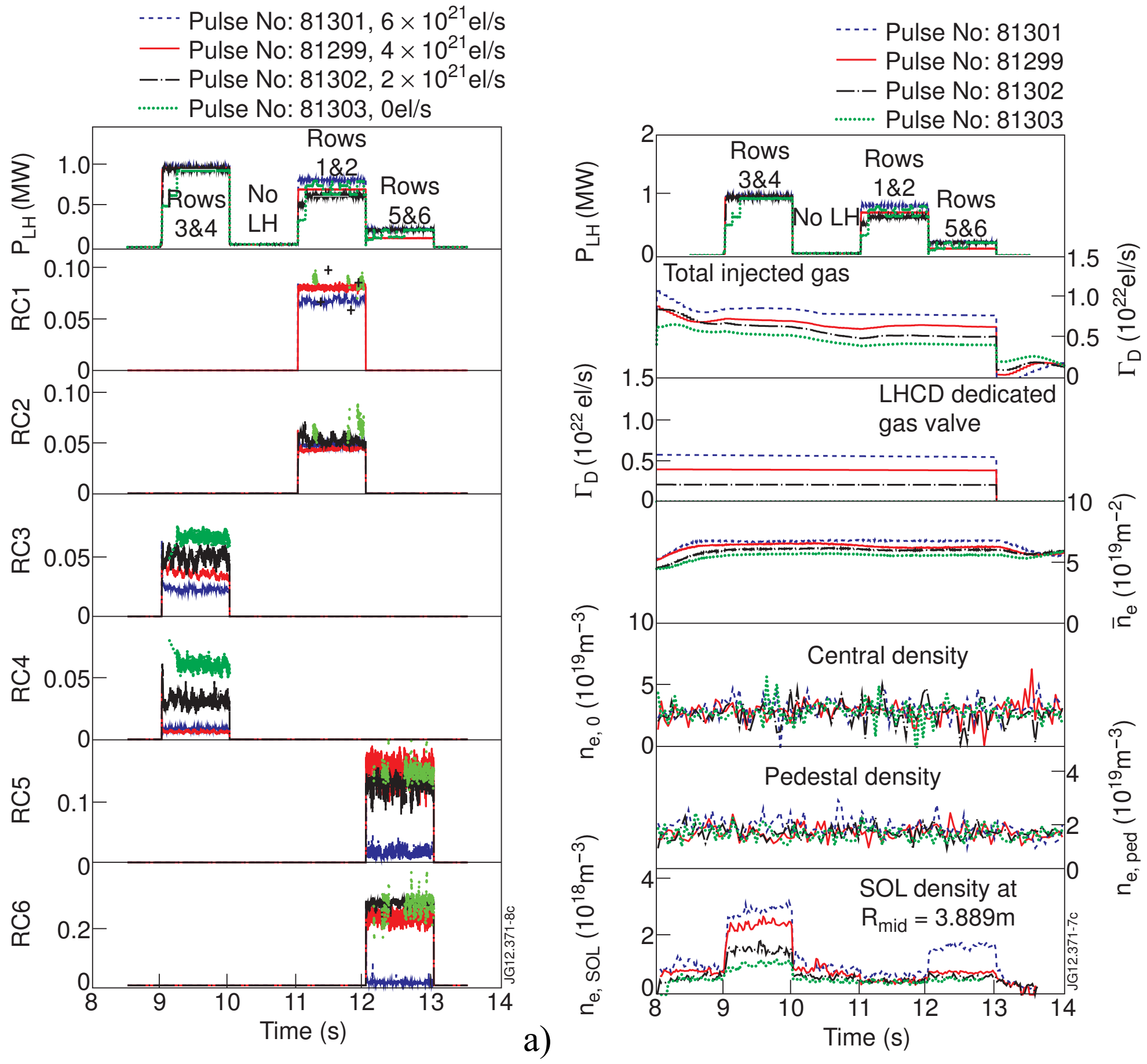

**Figure 4:** (a) LH RCs on rows 1 to 6 for four values of gas puff rate (provided in the legend) from dedicated gas injection module. Rows 3&4, 1&2 and 5&6 were pulsed consecutively as gap with no LH power was left in between. In all four pulses the launcher was aligned to the nPL, $l_{pos}$=0.000m, and same configurations and separatrix to PL distance, $r_{ROG}$=0.042m, was used. The RCs when klystrons are tripping are shown by symbols, black + for #61302, row1 and green dots for #61303, rows1, 2, 5, and 6. (b) Time traces of $D_2$ gas injection, total and from dedicated valve, electron density, line integrated, central and at the pedestal (normalised poloidal flux $\Psi_N$=0.9), and SOL density at midplane position of PL, $R_{mid}$=3.889m, by Li-beam.

## SOL density measurements.

The JET Li-beam diagnostic was substantially upgraded [32] during ILW installation work. Measurements were used to assess the Scrape-off Layer (SOL) density modifications due to LH. For plasmas with 2.7T/2.45MA or 2.2T/2MA the middle of the LH grill, i.e. rows 3&4, is magnetically connected to the diagnostic, while the top, rows 1&2, and the bottom of the launcher, rows 5&6, are not connected. For the gas scan pulses discussed in the previous Section, Li-beam measurements were performed and time traces of the power waveforms, gas puff rates, line integrated, central, pedestal and SOL densities are shown in figure 4b. The bottom graph provides the density in front of the PL with a midplane position of about $R_{mid}$≈3.889m, which is about 0.04m outside the separatrix. Density profiles for #81299 with gas rate from dedicated valve of $4\times10^{21}$el/s at different time slices are also shown in figure 5a. For the time slice at 9.85s magnetically connected rows 3&4 were pulsed and density increases between $R_{mid}$≈3.87m and 3.89m. At 10.09s, heating was turned off while at 10.85s and 11.09s the top, 1&2, and the bottom, 5&6, rows were pulsed respectively. As these parts of the grill are not connected to the Li beam diagnostic, the density stays unchanged. The observed effects are poloidally inhomogeneous and separated in a sense that only the flux tube in front of the powered sections of the grill show changes in density as it can be seen from time traces at the bottom of figure 4b as well. From the time traces in figure 4b and the profiles in figure 5a it is clear that the LH power modifies the SOL density profile in front of the powered sections of the launcher only. The modifications are local, i.e. when rows 1&2 are energised no changes to SOL density in the flux tube connected to rows 3&4 is seen as the time traces at the bottom of figure 4b remain flat in time intervals 11s-12s and 12s-13s. At a maximum gas injection rate row 5&6 can also affect the density in front of rows 3&4. As a result of these changes to the SOL density the LH wave coupling is improved, also shown by RCs trends in figure 4a. Individual profiles with rows 3&4 powered at ≈9.5s and no LH power at ≈10.5s for different rates of gas injection are shown in figure 5b. It was found that the increase in SOL density is ‘hump’ shaped and scales with gas injection rate. An interesting observation is that even without gas from a dedicated gas injection valve, the LH power affects the SOL density, as shown by the green profiles in figure 5b.

In addition, SOL density changes were studied as a function of LH power and launcher position. The SOL density increases with increasing the coupled RF power, figure 5c. In this example LH power was ramped-up (figure 5c inset graph) while the plasma parameters and the configuration were not changed indicating that processes involved are acting on a very fast time scale. It is worth noting that although negligible the effect is present even at the lowest applied power of 0.42MW. The increase of density in the hump was more dramatic, ≈70%, when the power was increased from 0.42MW to 0.87MW and very small, only ≈%10, for changes from 1.56MW to 2.02MW, figure 5c. Possible explanation of this observation could be that at higher power most of the injected gas is ionised and the amount of available neutrals is smaller hence lower ionisation rate and smaller increase in SOL density.

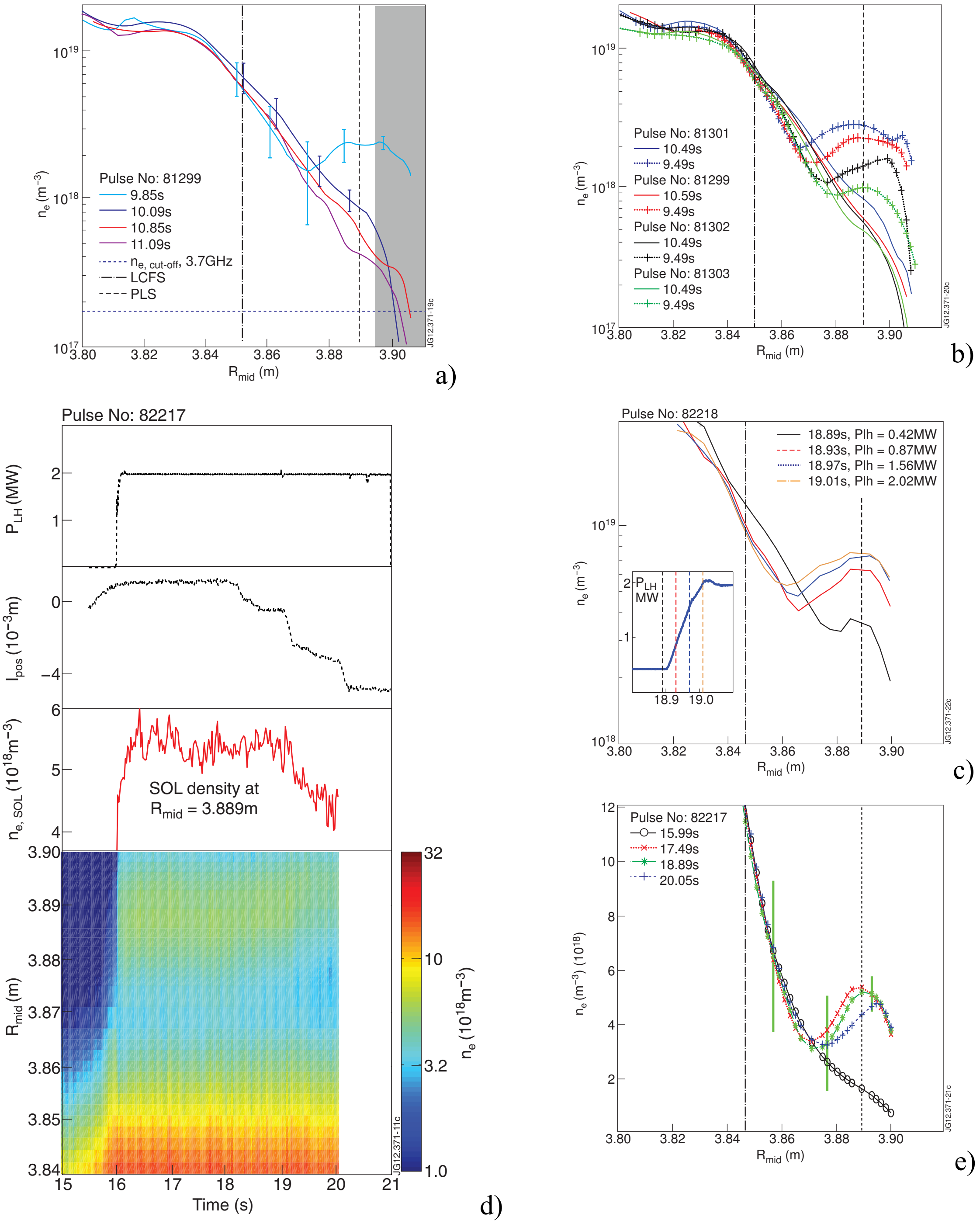

**Figure 5:** Li beam profiles for 2.7T/2.45MA JET pulse #81299 with gas injection rate from dedicated gas valve of $4\times10^{21}$ el/s and rows 3&4 magnetically connected to the diagnostic (a). Time traces of the power waveforms are provided in figure 4(a) as well. Private SOL region of Li-beam diagnostic where measured density is not directly affected by LH power is shadowed in (a) while separatrix (dash-dotted line) and poloidal limiter (dashed line) projections at the midplane

radius are shown for all radial profiles. SOL profiles with ('+' symbols) and without (solid lines) LH power for different gas injection rates, see also figure 4(b), are provided in (b). The impact of the microwave power can be seen in (c) where four SOL density profiles during LH power ramp-up (inset graph) are shown. Effect of the launcher position is shown in (d) where 2MW of LH power was applied in #82217 and launcher retracted back as shown by $l_{pos}$ trends. The density at the limiter ($R_{mid}$ =3.889m) and colorplots versus time and $R_{mid}$ are provided as well. SOL profiles at 15.99s (no LH power), 17.49s ($l_{pos}$≈+0.001m), 18.89s ($l_{pos}$≈0.000m) and 20.05s ($l_{pos}$≈-0.003m) are given in (e).

The behaviour of local density changes while retracting the launcher backward is shown in figure 5d, e. The evolving SOL density profile before LH power was turned on at 16s is due to changes in the configuration and the gas injection rates needed to form target plasma for good LH coupling. Once this is achieved shortly (at about 15.9s) before the LH pulse, the plasma density and shape were essentially kept unchanged during the whole heating phase. It can be concluded that the hump is moving with the launcher thus forming a big dip region when the latter is at its farthest position. At the same time the maximum of the hump decreases insignificantly (figure 5e). This is a possible indication that the processes involved in density modifications are local, i.e. occurring just in front of the launcher where the maximum of the hump is. The decrease in density is better pronounced after 20s when the launcher is retracted behind the nPL with $l_{pos}$≈-0.003m (see time traces in figure 5d). This observation is qualitatively consistent with changes in the scrape-off lengths of the flux tubes in front of the launcher, when latter is retracted behind the limiter. The colorplot on the bottom of figure 5d also shows that the density changes and in particular the formation of a hump happen instantly after LH power is turned on.

## EDGE2D modelling.

Local density modifications, and in particular plasma density rise in front of the launcher, were observed and modelled [33] with the two dimensional code EDGE2D [34] in the old JET conditions. The early modelling work [33] does not include local LH power dissipation, which can cause stronger density enhancement and even appearance of a hump (figure 5). As no reasonable modifications to the SOL transport were able to reproduce this feature, the EDGE2D code was modified [35], [50] to take into account the heating of the SOL by a fraction of the LH power parasitically dissipated in front of the grill [36]. Because of the lack of a complete theory to explain the latter, the amount of LH power lost in the SOL in JET can be estimated [35] using the experimental scaling law obtained in Tore Supra [51]. The power loss is then presumably transformed into electron heating in the SOL and it can be accounted for in the EDGE2D code by modifying the value of the energy source to electrons accordingly. Details on this approach are given in [35] and the very same methodology is used here. In the following a brief discussion on the possible mechanisms for LH power dissipation in the SOL is presented.

In general, the SOL density modifications with LH power observed in this experiment can be discussed in the context of density transport equation. The appearance of a hump can be either due to local changes in the ionisation source or changes in the SOL transport or combination of both. Changes in the SOL turbulence, induced by $CD_4$ gas puff from dedicated gas valve could affect the SOL transport [33]; however, in our case $D_2$ gas was used in ILW conditions and it seems effect is present even without gas puff (see green lines in figure 5b).

It is also possible that parasitically absorbed LH power in the SOL generates fast electron populations which can significantly influence the SOL potentials [45] and induce **E**x**B** drifts, which might impact on the SOL transport as discussed in [46]. This idea is formulated in [45], [52] where it is shown that as the LH power is applied the sheath potentials in the vicinity of the LH launcher are altered resulting in **ExB** flows which in turn lead to density modifications near the antenna. This assumption is however difficult to be investigated in our case as no proper measurements and mapping of the electric fields in front of the launcher can be made. Initial assessment of the role of the drifts [53] indicated that this effect is small at JET. Similarly, also the parallel ponderomotive force expelling plasma from the front of the grill, gives only a minor correction to the local density balance.

Ideally, the spectrum of the launched LH wave will be peaked at around $N_{\|}\approx 1.8$ and the RF power can not be absorbed in the cold SOL by the bulk electrons. Although a number of mechanisms which can up-shift the launched $N_{\|}$ spectrum exist neither of them seems applicable in our case. Strong scattering of launched LH wave by parametric decay instabilities (PDI) may be responsible for power losses near the plasma periphery if the ratio of the launched wave's to lower hybrid frequencies satisfies the condition $\omega_0/\omega_{LH} \leq 2$ as noted in [47]. For the case presented in figure 5c it was estimated that $\omega_0/\omega_{LH} > 6$ in which case there should be no issue with PDI. Strong wave scattering due to density fluctuations near the plasma edge can be also considered as a possible cause of LH wave spectrum up-shift. Calculations showed that only about 0.04m inside the separatrix is accessible for LH waves launched with $N_{\|}\approx 1.8$. Therefore, a number of reflections can be expected, which is a necessary condition for strong scattering of the LH wave by density perturbations. It is worth mentioning, however, that the plasma core is accessible for $N_{\|}>2$ so once spectrum of the LH wave up-shifts to this value the wave is expected to penetrate fully into plasma. Therefore, it can be concluded that the accessibility and scattering on density fluctuations should not affect significantly the launched LH wave spectrum.

Taking into account the fact that the LH launcher is not ideal and a fraction of the wave power is launched at relatively high-$N_{||}$, i.e. $| N_{||} | > 30$, a small population of energetic electrons can be generated in front of the launcher as shown in recent modelling works [35], [36]. In addition power absorption via collisional damping [48], [49], [11] could account for the reduction in power absorbed in the confined plasma.

Furthermore the observed density changes were investigated by means of EDGE2D code. Geometrical and SOL related parameters were all input into the code in agreement with the available experimental data. In the simulations the gas was injected from poloidal locations corresponding to the dedicated gas injection valve location. It was assumed that density modifications in the SOL with LH power are due to enhancement in the ionisation rates caused by a local LH heating of the SOL. The RF power losses were assessed according to the scaling used in [51] and were subsequently introduced in the code as a fixed source of heating in electron energy balance. The modelling results in which SOL heating about 2cm in front of the launcher is assumed are shown in figure 6.

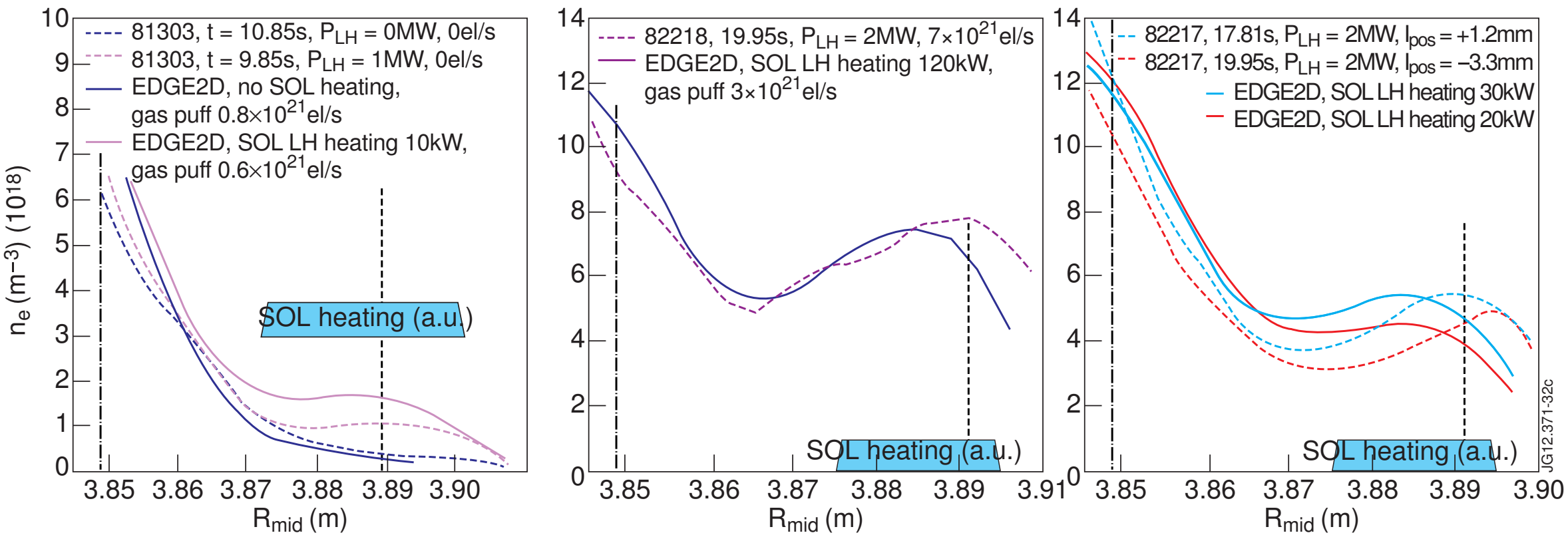

a) b) c)

**Figure 6:** Measured Li-beam SOL density profiles (thick dashed lines) without (blue) and with (magenta) LH power for shot #81303 (a). EDGE2D modelling results are provided (thin solid lines) assuming zero (blue) or 30kW (magenta) of parasitic LH heating in a.u. between $R_{mid}$=3.87m and $R_{mid}$=3.90m (cyan rectangle). LH power and the gas injection rates – as used in the model and from dedicated valve – are given in the legend. Experimental data (thick dashed line) and EDGE2D (thin solid line) simulations for shot #82218 with gas injection rate from dedicated vale of $7\times10^{21}$ el/s are shown in (b). Li-beam data and EDGE2D simulations for shot #82217 when 2MW of LH power was applied while launcher placed at different positions are shown in (c). Vertical dash-dotted and dashed lines show the position of the separatrix and the poloidal limiter at the midplane.

The simulations can predict reasonably well the changes in SOL density profiles with LH power, figure 6a, gas injection rate, figure 6a and 6b, and launcher position, figure 6c. In the case when no gas from dedicated valve was used the code was first tested and found in good agreement with measurements for $P_{LH}$=0 (see blue lines in figure 6a).

The changes in SOL profile with LH power were best simulated assuming about 10kW of LH power absorbed in the SOL (magenta lines in figure 6a). Although no power scan was performed in the study it should be noted that the increase in density in the modelling scales with the estimated LH power losses in the same way as it is observed in the experiment with respect to the launched power. A good illustration of this feature of the model can be also found in [35] (figure 7 in [35]) in which particular case for dissipated powers of 10 and 50kW the resulting density changes were shown to be consistent with the measurements for launched powers of 0.4 and 1.6 MW.

The calculations shown in figure 6 are more consistent with the experimental results in the cases when gas from dedicated valve was used; however, in the simulations about twice smaller gas puff rate was needed to reproduce the changes in SOL density with LH power (see figure 6b). An increase in dissipated power was needed in order to reproduce the density changes when launcher is moved forward in front of the limiter (figure 6c). From these simulations it can be concluded that only a few percent (maximum of about 5-6%) of the launched power is absorbed by the SOL plasma in the magnetic flux tubes in front of the LH grill. Further improvement of the SOL modelling will account for possible changes in SOL transport with LH power in an attempt to achieve even better fit to the measured profiles. For the purpose of the paper, however, it is important to note that the observed density changes can be reproduced assuming small amount of RF power is absorbed in the SOL.

## ARCS AND HOTSPOTS

Issues related to possible harmful interaction between LH waves and plasma, in particular in the SOL region, were monitored via the available viewing systems. In addition to the new cameras used for protecting first wall components of JET, dedicated diagnostics have been installed to protect against, and study, the parasitic loss of LH power in the SOL (respectively, pyrometers viewing the LH launcher and the nearest poloidal limiter, and a dedicated infrared (IR) camera). An unfolded view of JET interior showing camera views is provided in figure 7. Projections of magnetic field lines covering the LH launcher boundaries from top to bottom in a typical 2.4T/2MA conditioning pulse are provided as well. Shown are the views of the cameras used in the study, namely, LH camera, wide angle IR and divertor camera. LH camera has visible CCD and IR imaging system. The approximate positions of the hotspots discussed in the study are indicated as well.

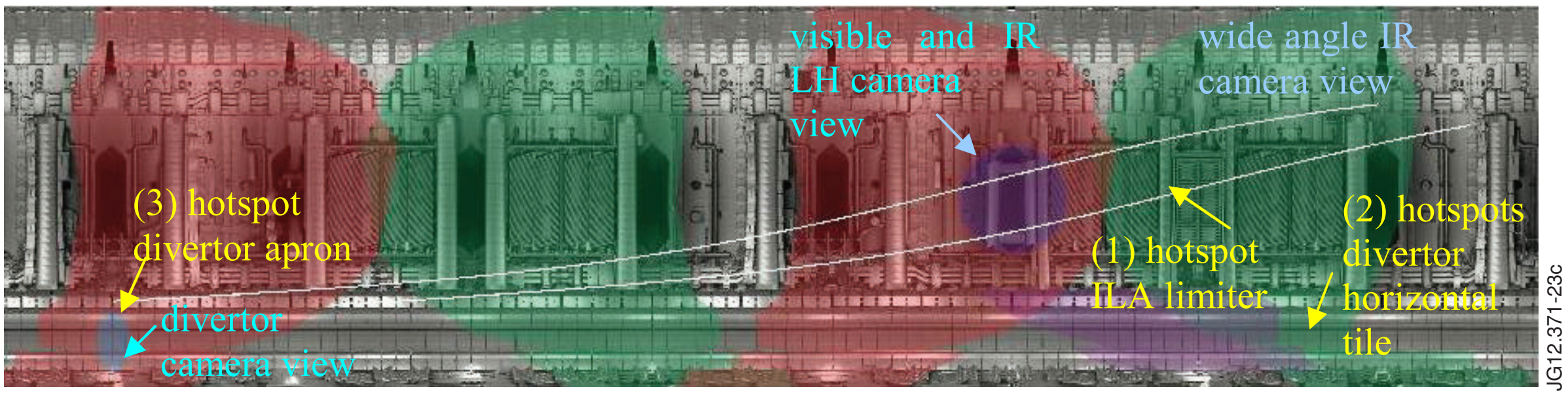

**Figure 7:** A sketch of JET cameras view with two magnetic field lines covering the front of the LH launcher for a typical 2.4T/2MA JET pulse #81382, 18.08s. Hotspots positions are provided by yellow arrows and labels, while cameras' views are indicated by cyan arrows and labels.

Arcs in front of the LH grill can be monitored by the visible LH camera, figure 7, while hot spots can be seen by IR camera. Examples of hotspot observation presented here are on (1) the ICRH ITER Like Antenna (ILA) antenna left side limiter seen by wide angle IR camera; (2) on horizontal tile of the outer divertor and (3) on the top of divertor apron as seen by the divertor camera. All these places are indicated by arrows in figure 7.

**Observation of arcs via a dedicated visible camera.**

Amongst the number of protection issues which were addressed during the LH operation with ILW at JET, arcs were thought to be potentially the most dangerous events. Arcing is dangerous not only because it causes damage to the launcher but also due to the associated large impurity influx of *Fe*, which sometimes causes plasma disruption. Arcs are now monitored by a new dedicated visible camera, figure 7, viewing the whole LH grill with sufficient resolution to identify on which part of the LH grill the arc is taking place.

For the first time at JET the dynamics of arc development have been observed (figure 8). It has been seen that if a localised arc is not extinguished fast enough by the existing protection systems, (based on the reflected power imbalance and on the impurity radiation) it can propagate along the grill mouth thus covering a much broader area, which might result in substantial damage to the launcher. In rare cases events of huge arcs are followed by plasma disruption.

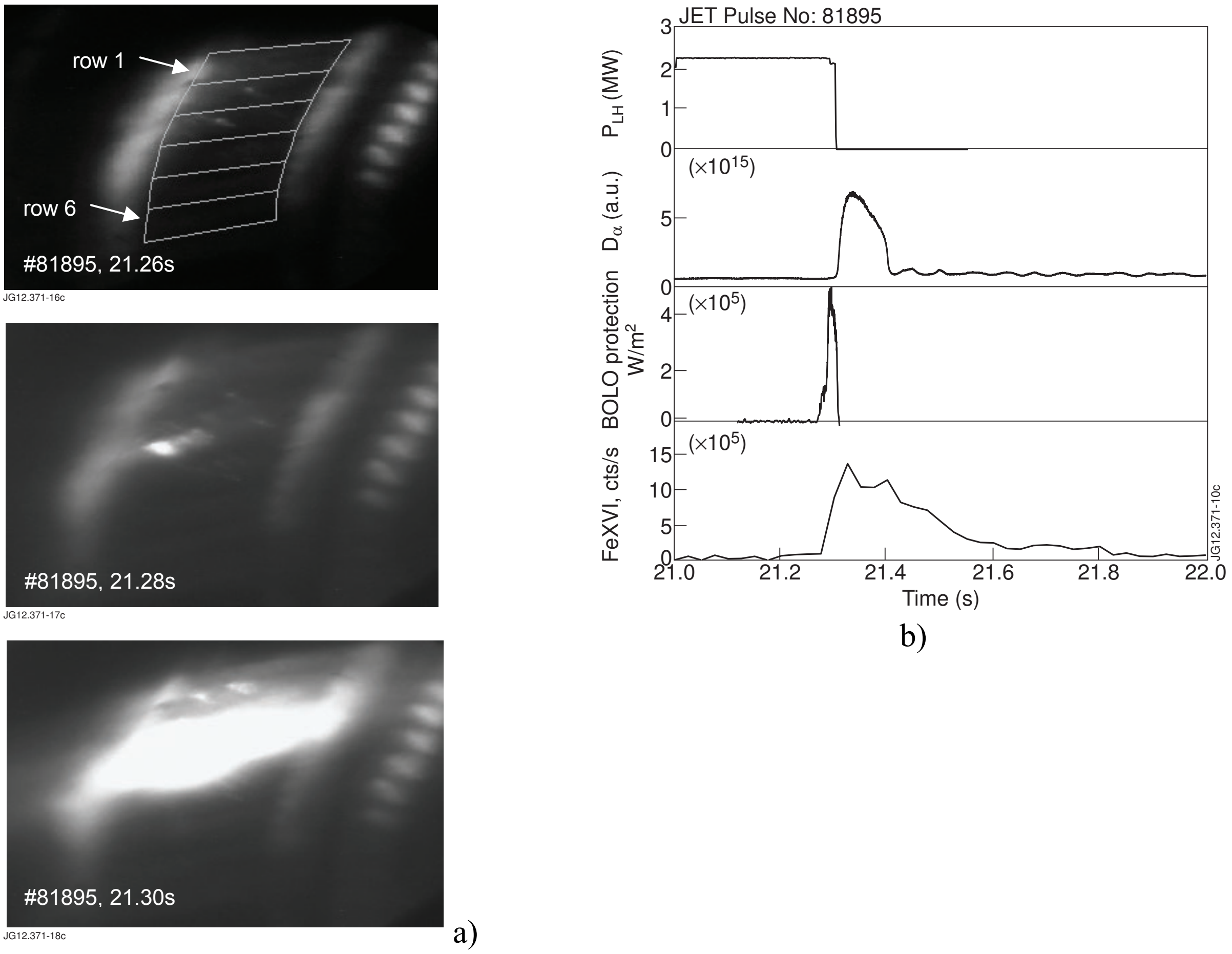

**Figure 8:** Arc seen on the visible LH camera in three consecutive frames, #81895, 21.26s, 21.28s and 21.30s, showing how an arc develops into a flare in front of rows 2&3. On the right, the time traces of the LH power, $D_\alpha$ signal as well as the signals from bolometer and FeXVI line used in arc protection based on impurity radiation are shown.

## Heat loads on the launcher via IR camera and hotspots.

Hotspots on a PFC magnetically connected to the LH launcher were observed before on JET [37], [38], [39] and on other machines [40], [41], [42], [43]. In our study we refer to all areas seen on the IR cameras which are brighter than normal as hotspots. Clarification is needed, however, as LH waves can cause excessive heating of PFC either (*i*) by parasitically accelerated energetic electrons in SOL, which follow magnetic field lines and thus intercept on different in-vessel objects or (*ii*) by contributing to the power input and exhaust. In the former case hotspots are localised and magnetically connected to the LH launcher. In the second case brighter areas associated with either debris or surface layers can be seen usually in the divertor region during LH power phase, but not necessarily in regions on the same flux tubes as the launcher. The hotspots due to fast electrons are directly related to LH wave and SOL interaction and are thought to be more important regarding inner wall protection. While on the carbon-based machines allowable temperatures were much higher, surface temperatures above 950$^o$C were not allowed for *Be* PFC, and so special attention has been paid to the hotspots during LH in conditions with ILW. In general, hotspots due to LH were noticed during restart sessions; however, up to now no detrimental temperature increase has been observed for pulses up to 2.5MW of RF power and up to 5s duration.

When the LH power was increased above 2MW and with launcher placed between $l_{pos}$= -0.005m and +0.005m a hotspot in the middle of the ILA limiter can be clearly seen by the wide angle IR camera. The view of the camera and the approximate position of the hotspot, labelled as (1), are shown in figures 7 and 9. Its temperature increase was correlated with the LH power and it reached about 340$^{o}$C (figure 9) which is well below the safety limit of 950$^{o}$C for the limiter *Be* tiles. The large time constant characterizing the temperature increase/drop after LH switches on/off is an indication that a bulky limiter component has been warmed rather than a surface effect taking place. It was noted that the hotspot on ILA limiter moves downward with increasing equilibrium magnetic field, $B_t$, which proves that it is related to parasitical heating of a magnetically-connected PFC by an energetic electron beam confined within a magnetic flux tubes.

There is also an increase in temperature on the horizontal tiles on the bottom part of the outer divertor (labelled as (2) in figure 9), which is correlated with LH power. Preliminary investigation shows that no part of the LH launcher is magnetically connected to this area as shown in figure 7. It was concluded that the observed bright spots are actually debris (usually observed at the bottom of the divertor) and the increase of their temperature is due to the combination of plasma configuration with strike point at the corner used in LH pulses and the increase of the exhaust power. No relevant temperature measurement for these hotspots is available as the IR camera was not calibrated for *W* coated tiles and the existence of debris make the analysis and the interpretation of the results more difficult. Close examination of ICRH-only pulses at similar corner configuration (see figure 3b) shows the same trends and the same relative increase in temperature with ICRH power. The conclusion is then that the observed bright areas are purely due to heating of debris and cannot be related to parasitic loss of LH power but are due to the combination of configuration change and exhaust power increase associated with LH.

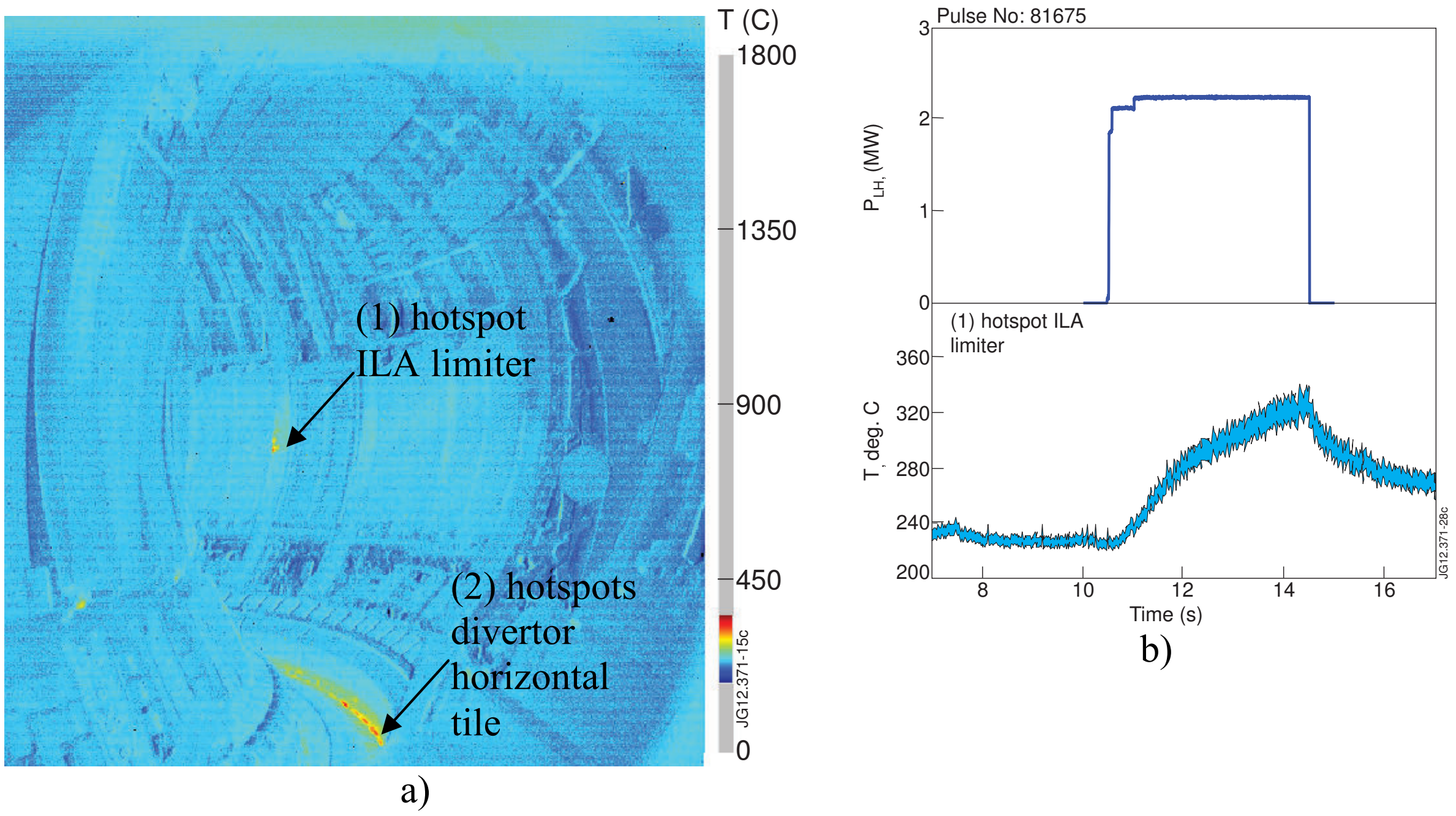

**Figure 9:** IR image of the plasma interior during a 2.3MW LH JET pulse #81675, 14.42s (a) and time traces of the LH power and temperature range of hotspot (1) on ILA limiter (b). No relevant temperature measurement for hotspots (2) is available as the IR camera was not calibrated for *W* coated tiles and the debris make the analysis more complicated.

In some cases the LH was misleadingly thought to cause hotspots. This is mostly seen on the divertor camera shown in bottom left corner of figure 7. More detailed investigation [19] has shown that the fast increase and drop of the temperature with LH power observed in this case is an indication that probably a thin surface layer has been heated or other surface effect is taking place. Indeed, warming bulky PFC would result in much slower temperature variations as shown by time traces in figure 9b.

Images from an IR camera were used for the first time on JET [24] to assess the distribution of the heat on the launcher mouth during LH-only operation. Provided that the launcher is melted predominantly on the left side and also at the top left corner [44] there is no clear indication that the hottest areas of the grill are the most damaged multijunctions.

## SUMMARY AND CONCLUSIONS

Overall, a relatively trouble-free operation of the LH system in the new ILW for up to 2.5MW of coupled microwave power in L-mode plasma for time duration of up to 5s was achieved. The figure of 2.5MW/5s was set as a protection limit for initial operation with ILW and there is no indication that the power cannot be increased further.

No significant impact on *Be* radiation with LH power was observed. No *W* accumulation with LH-only heating is reported.

A general observation from the initial LH operation with ILW is that the LH coupling is not degraded with installation of ILW in conditions with $l_{pos}$≥0.000m. Similarity studies shown in figure 3 indicated that the Reflection Coefficients on rows 1 to 4 have similar values during conditioning pulses with the old and the new wall. RCs on rows 5 and 6 are higher in this example but sufficient $D_2$ injection from dedicated gas valve improves the coupling on the bottom part of the launcher as well.

Improved Li-beam measurements allowed for SOL density measurements and first systematic study of the impact of LH on the SOL parameters. Gas injection rates, plasma density and configuration and launcher position were all scanned in optimising the coupling, while the relevant changes to the SOL parameters were documented by Li-beam measurements. The application of LH power leads to local increase in density in front of the powered sections of the grill. SOL density modifications have a typical ‘hump’ shape and scale with gas injection rate and LH power. Experimental results are supported by EDGE2D modelling.

First observation of arcs in front of LH grill is reported. Arcs are not always stopped by existing protection and when not extinguished in time they seem to propagate along the row. A proposal to implement a new real-time protection acting only on the arcing klystrons is being considered as part of JET protection system and a project on its implementation has been started. It is believed that with this protection in place arcs can be stopped quickly enough to avoid the potentially dangerous consequences of damage to the launcher and plasma disruption.

Hotspots were observed but not found to be causing problems for powers up to 2.5MW and pulse durations of 5sec.

## ACKNOWLEDGEMENT

This work, supported by the European Communities under the contract of Association between EURATOM and CCFE, was carried out within the framework of the European Fusion Development Agreement. The views and opinions expressed herein do not necessarily reflect those of the European Commission. This work was also part-funded by the RCUK Energy Programme under grant EP/I501045

One of the authors (V. P.) acknowledges support by the Czech Science Foundation Project 205/10/2055, and of MSMT CR Grant Project ID LG11018.

Authors would like to thank S. Devaux, G. Corrigan, P. da Silva Aresta Belo and D. Harting for help with IR camera data and EDGE2D code.